\documentclass[doublecol]{epl2} 
\usepackage{color}
\usepackage{verbatim}
\usepackage{hyperref}
\usepackage{epsf}
\usepackage{graphics}
\usepackage{graphicx}
\usepackage{dcolumn}
\usepackage{bm}
\usepackage{amsmath}
\usepackage{amssymb}

\title{Identifying spin-triplet pairing in spin-orbit coupled multi-band 
  superconductors}

\author{Christoph M. Puetter\inst{1} \and Hae-Young Kee
  \inst{1,2}\thanks{E-mail: \email{hykee@physics.utoronto.ca}}}
\shortauthor{Christoph M. Puetter and Hae-Young Kee}

\institute{                    
  \inst{1} Department of Physics, University of Toronto, Toronto,
  Ontario M5S 1A7, Canada \\
  \inst{2} Canadian Institute for Advanced Research, 
  Quantum Materials Program, Toronto, Ontario M5G 1Z8, Canada
}

\abstract{
  We investigate the combined effect of Hund's and spin-orbit (SO) coupling 
  on superconductivity in multi-orbital systems.  
  Hund's interaction leads to orbital-singlet spin-triplet superconductivity,
  where the Cooper pair wave function is antisymmetric under the exchange of 
  two orbitals.
  We identify three $d$-vectors describing even-parity orbital-singlet 
 spin-triplet pairings among $\text{t}_{\text{2g}}$-orbitals, and
 find that the three 
  $d$-vectors are mutually orthogonal to each other.
  SO coupling further assists pair formation, pins the orientation 
  of the $d$-vector triad, and induces
  spin-singlet pairings with a relative phase difference of $\pi/2$.
  In the band basis the pseudospin $d$-vectors are aligned along 
  the $z$-axis and correspond to momentum-dependent inter- and 
  intra-band pairings.
  We discuss quasiparticle dispersion, 
  magnetic response, collective modes, and experimental consequences
  in light of the superconductor Sr$_{2}$RuO$_{4}$.
}

\begin{document}

\maketitle

\section{Introduction}
Since its inception, standard Bardeen-Cooper-Schrieffer (BCS) 
theory has been considered 
a classic example for a collective phase emerging from 
quantum many body effects.
However, the discovery of unconventional superconducting phases 
near antiferromagnetic order in heavy fermion compounds
\cite{Grewe91North,Sigrist91RMP}, organic materials \cite{Powell10arXiv},
and, most recently, Fe-pnictides \cite{Kamihara08JACS} 
have exposed the limits of a single-band BCS formulation.
The origin and nature of superconductivity in complex materials
where multiple bands cross the Fermi level
therefore remains a field of active research, harbouring intriguing 
challenges and mysteries.

In particular, when the electronic structure near the Fermi energy
is composed of different orbitals and spins mixed via spin-orbit (SO) coupling, 
a pairing symmetry analysis could be non-trivial.
For example, a local microscopic interaction 
such as Hund's coupling may naturally favour inter-orbital spin-triplet
pairing between electrons.
However, when orbital and spin fluctuations are significant 
due to inter-orbital hopping and SO interaction, 
pairing in definite orbital and spin channels 
(\emph{e.g.}, spin-singlet or -triplet pairing between electron in orbitals
$a$ and $b$)
is not well defined. 
Equivalently, from a Bloch band perspective, where
the kinetic Hamiltonian including SO effects is diagonal, 
the decoupling of the microscopic interaction
effectively leads to intra- and inter-band pairing
with pseudospin-singlet and/or -triplet character.

Below we present a systematic study of how  
SO and Hund's couplings jointly give rise to superconductivity 
in $t_{\text{2g}}$ (\emph{i.e.}, d$_{yz}$, d$_{xz}$, and d$_{xy}$) orbital systems.
Our findings may apply to a number of multi-orbital 
$d$-subshell superconductors. 
To be specific we base our quantitative considerations on the
proposed chiral spin-triplet superconductor Sr$_{2}$RuO$_{4}$.
Here, despite intense investigation for more than a decade, 
a clear picture for the pairing symmetry, the pairing mechanism
and the relevant bands involved that is consistent with all  
experimental observations has not yet emerged
\cite{Mackenzie03RMP,Kallin09JPhysCondensMatt}.
 
The paper is organized as follows.
In the second section we discuss Cooper pairing in multi-orbital systems.
We find that superconductivity from local Hund's exchange can
naturally be characterized by three mutually orthogonal d-vectors 
each describing inter-orbital \emph{even-parity spin-triplet} pairing. 
We then show how SO coupling pins the orientation of the 
d-vector triad and induces and enhances pairing via coupling
to spin-singlet pairing order parameters
with a fixed relative phase difference of $\pi/2$.
In the third section, 
we map these local pairing order parameters, 
defined in an orbital and spin 
basis, to inter- and intra-band pairing in the Bloch band basis. 
Pairing in the Bloch bands has a strong momentum dependence and the
magnitude and direction of the d-vectors depend on the orbital composition 
at each ${\bf k}$-point. 
In the fourth section, we present the complete self-consistent 
mean-field (MF) results involving 9 complex order parameters using 
band structure parameters that reproduce the Fermi surface (FS) reported on 
Sr$_2$RuO$_4$. 
In addition, the resulting anisotropic quasiparticle (QP)
dispersion, the magnetic response and 
the critical pairing strengths in the presence of SO coupling are considered. 
We summarize our findings and discuss the relevance 
for SO-coupled d-orbital superconductors such as Sr$_2$RuO$_4$ in the last 
section.

\section{Pairing in SO coupled t$_{\text{2g}}$ systems via Hund's interaction}
For multi-orbital 3d-subshell systems such as 
the Fe-pnictides, it was recognized that Hund's coupling 
(interaction strength denoted by $J$) is as important as 
on-site Coulomb repulsion ($U$) \cite{Lee08PRB,Yang09PRB},
while SO coupling ($2 \lambda$) is relatively weak \cite{Fazekas99World}.
In contrast, recent x-ray measurements on 5d transition metal compounds
such as Ir-based oxide materials found that the SO interaction of  
0.6 eV is roughly comparable to the on-site Coulomb energy \cite{Kuriyama10APL},
suggesting that SO interaction is larger than Hund's exchange
(since $J < U$).
Given that the effective pairing interaction in 
the spin-triplet channel arising from Hund's coupling
and inter-orbital Hubbard repulsion ($V=U-2J$)
scales as $V-J=U-3J$ (see below), we therefore expect that for 
4d-subshell materials such as 
Sr$_{2}$RuO$_{4}$ both SO and spin-triplet pairing interactions
are intermediate in strength and of similar magnitude
\cite{Liebsch00PRL,Mravlje11PRL,Pavarini06PRB,Malvestuto11PRB,Pchelkina07PRB,Haverkort08PRL,Rozbicki11JPhysCondensMatter}.
Since neither interaction is negligible nor dominant, we
treat both on an equal footing in the present study.

While on-site Hund's and further neighbor exchange 
interactions have been recognized to be important 
for spin-triplet pairing \cite{Lee08PRB,Ng00EPL,Spalek01PRB,Han04PRB,Dai08PRL},
the combined effect of SO and Hund's couplings on inter-orbital spin-triplet
pairing has not been investigated in t$_{\text{2g}}$-orbital systems.
To understand superconductivity in SO coupled t$_{\text{2g}}$-orbital systems, 
we consider a generic Hamiltonian 
$H = H_{\text{kin}} + H_{\text{SO}} + H_{\text{int}}$
consisting of kinetic, SO, and local Kanamori interaction terms.
In this section we leave the kinetic Hamiltonian $H_{\text{kin}}$
unspecified and focus on the pairing properties 
arising from the interplay of the atomic SO coupling 
$H_{\text{SO}}= 2 \lambda \sum_{i} {\bf L}_{i} \cdot {\bf S}_{i}$
and the local interaction, which,
projected on the t$_{\text{2g}}$ orbitals, are given by 
\begin{eqnarray}
  \label{eq:Hint}
   H_{\text{SO}} &=& i \lambda 
  \sum_{i} \sum_{a b l}
  \epsilon_{a b l} c^{a \dagger}_{i \sigma} c^{b}_{i \sigma'}
  \hat{\sigma}^{l}_{\sigma \sigma'}, \\
  H_{\text{int}} &=&  \frac{U}{2} \sum_{i, a}   
  c^{a \dagger}_{i \sigma}  c^{a \dagger}_{i \sigma'}
  c^{a}_{i \sigma'}  c^{a}_{i \sigma}
  + \frac{V}{2} \sum_{i, a \neq b}  
  c^{a \dagger}_{i \sigma}  c^{b \dagger}_{i \sigma'}
  c^{b}_{i \sigma'}  c^{a}_{i \sigma}  \nonumber \\
  && + \frac{J}{2} \sum_{i, a \neq b} 
  c^{a \dagger}_{i \sigma}  c^{b \dagger}_{i \sigma'}
  c^{a}_{i \sigma'}  c^{b}_{i \sigma} 
  + \frac{J'}{2} \sum_{i, a \neq b} 
  c^{a \dagger}_{i \sigma}  c^{a \dagger}_{i \sigma'}
  c^{b}_{i \sigma'}  c^{b}_{i \sigma}. \nonumber \\
  \label{eq:Hint2} 
\end{eqnarray}
Here and in the following, summation over repeated  
spin indices \mbox{$\sigma, \sigma'=\uparrow, \downarrow$} is implied
while the indices $a, b \in  \{yz, xz, xy\}$  
belong to an ordered set of $t_{\text{2g}}$-orbitals.
Furthermore, $\hat{\sigma}^{l}$ stands for 
Pauli matrices, $c^{a \dagger}_{i \sigma}$ creates an electron on site $i$ in 
orbital $a$ with spin $\sigma$, and $\epsilon_{a b l}$
denotes the totally antisymmetric rank-3 tensor.
For transparency we have also introduced separate interaction strengths for 
Hund's coupling ($J$) and pair hopping ($J'$), although
$J = J'$ at the atomic level.

\begin{figure}[t!]
  \centering
  \includegraphics*[width=0.8\linewidth, clip]{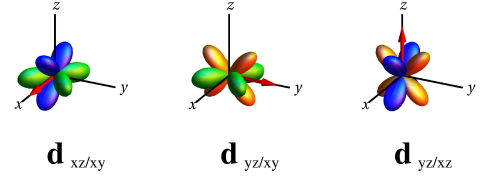}
  \caption{(Color online) The orbital-singlet spin-triplet 
    $d$-vectors form a triad whose orientation is pinned along
    $\hat{\bf x}$, $\hat{\bf y}$, and $\hat{\bf z}$
    (or $-\hat{\bf x}$, $-\hat{\bf y}$, and $-\hat{\bf z}$)
    in the presence of SO coupling. See main text for details.
    \label{fig:DVectors}}
\end{figure}

Let us apply a MF approach to study the particle-particle instabilities
of the microscopic interaction $H_{\text{int}}$ using the following 
zero momentum pairing channels 
\begin{eqnarray}
  \label{eq:MF1}
  \hat{\Delta}^{s}_{a/b} &=& \frac{1}{4 N} 
  \sum_{{\bf k}} [i \hat{\sigma}^{y}]_{\sigma \sigma'} 
  (c^{a}_{{\bf k} \sigma} c^{b}_{-{\bf k} \sigma'} 
  + c^{b}_{{\bf k} \sigma} c^{a}_{-{\bf k} \sigma'}), \\
  \hat{d}^{l}_{a/b} &=& \frac{1}{4 N}
  \sum_{{\bf k}} [i \hat{\sigma}^{y} \hat{\sigma}^{l}]_{\sigma \sigma'} 
      (c^{a}_{{\bf k} \sigma}  c^{b}_{-{\bf k} \sigma'} 
      - c^{b}_{{\bf k} \sigma}  c^{a}_{-{\bf k} \sigma'}) , 
      \label{eq:MF2}
\end{eqnarray}
where $N$ is the number of ${\bf k}$ points.
Here, 
$\Delta^{s}_{a/b} = \langle \hat{\Delta}^{s}_{a/b} \rangle$ 
$(= \Delta^{s}_{b/a})$ stands for intra- ($a = b$)
and inter-orbital ($a \neq b$) spin-singlet pairing, which is 
even under the exchange of orbital quantum numbers 
(\emph{i.e.} they form ``orbital triplets'').
The vector order parameter 
${\bf d}_{a/b} = (\langle \hat{d}^{x}_{a/b} \rangle, 
\langle \hat{d}^{y}_{a/b} \rangle, \langle \hat{d}^{z}_{a/b} \rangle)$
($= -{\bf d}_{b/a}$) 
on the other hand parametrizes inter-orbital ($a \neq b$)
spin-triplet pairing consistent with the usual $d$-vector notation
where  $i ({\bf d} \cdot {\bf \hat{\sigma}}) \hat{\sigma}^{y}$
describes the spin-triplet pairing gap
\cite{Leggett75RMP,Sigrist91RMP}.
Note that ${\bf d}_{a/b}$ is odd under orbital exchange,
which is characteristic of an ``orbital singlet'' (while ${\bf
  d}_{a/a} = 0$).
Note also that the above order parameters are all even under a 
parity transformation as they are locally defined;
this feature differs in particular from conventional 
odd-parity spin-triplet pairing where orbital degrees of freedom are absent.

Using the above pairing channels the interaction Hamiltonian takes the form 
\begin{multline}
  \label{eq:HMFPairing}
  H_{\text{int}} \rightarrow
   U N \sum_{a} \hat{\Delta}^{s \dagger}_{a/a} \hat{\Delta}^{s}_{a/a}
   + (V - J) N \sum_{a, b, l} 
  \hat{d}^{l \dagger}_{a/b} \hat{d}^{l}_{a/b}
  \\
  + J' N \sum_{a \neq b} 
  \hat{\Delta}^{s \dagger}_{a/a} \hat{\Delta}^{s}_{b/b} 
  + (V + J) N \sum_{a \neq b} 
  \hat{\Delta}^{s \dagger}_{a/b} \hat{\Delta}^{s}_{a/b},
\end{multline}
where it is clear that only Hund's coupling can give rise to an instability in 
a spin-triplet channel \cite{Spalek01PRB,Lee08PRB}.
We thus concentrate on the effective pairing interaction 
\begin{eqnarray}
  H'_{\text{int}} = 
  (U - 3 J) N \sum_{a, b, l} 
  \hat{d}^{l \dagger}_{a/b}
  \hat{d}^{l}_{a/b}
\end{eqnarray}
in the attractive regime $U/3 < J$ ($< U$).
In general, orbital-singlet spin-triplet pairing can also induce
spin-singlet pairing so that the remaining terms in Eq. (\ref{eq:HMFPairing}) 
would hamper spin-singlet pairing. 
However, we assume that their effect is negligible to keep the 
following self-consistent calculations feasible, and since the
induced spin-singlet pairing amplitudes are for the most part
smaller than the spin-triplet pairing amplitudes (see below).
For notational clarity we label in the following 
inter-orbital pairing only by the 
three combinations $a/b = xz/xy, yz/xy, yz/xz$.

To understand the effect of SO interaction, 
let us remark on pairing in the absence of SO coupling first.
In the case of the layered compound considered below (and for a rather
large parameter range) the three spin-triplet d-vectors 
${\bf d}_{xz/xy}$, ${\bf d}_{yz/xy}$, and ${\bf d}_{yz/xz}$ 
form a triad of mutually orthogonal vectors with an arbitrary orientation 
and chirality in spin space, and no relative complex phase difference
(hence preserving time reversal symmetry (TRS)). 
This can be understood by analyzing the Ginzburg-Landau (GL) free energy,
which without SO coupling is given by 
\begin{eqnarray}
  {\cal F} 
  &\sim& \sum_{\nu} \big[ A_{\nu} |{\bf d}_{\nu}|^{2} 
  + B^{(1)}_{\nu} ({\bf d}_{\nu} \cdot {\bf d}^{*}_{\nu})^{2} 
  + B^{(2)}_{\nu} |{\bf d}_{\nu} \cdot {\bf d}_{\nu}|^{2} \big]
  \nonumber \\
  &+& \sum_{\nu \neq \kappa} 
  \big[ C^{(1)}_{\nu \kappa} 
  ({\bf d}_{\nu} \cdot {\bf d}_{\nu}) 
  ({\bf d}_{\kappa} \cdot {\bf d}_{\kappa})^{*}
  + C^{(2)}_{\nu \kappa} 
  |{\bf d}_{\nu}|^{2} |{\bf d}_{\kappa}|^{2} \\
  &+& C^{(3)}_{\nu \kappa} 
  |{\bf d}_{\nu} \cdot {\bf d}_{\kappa}|^{2} 
  + C^{(4)}_{\nu \kappa} 
  |{\bf d}_{\nu} \cdot {\bf d}^{*}_{\kappa}|^{2} 
  + C^{(5)}_{\nu \kappa} 
  ({\bf d}_{\nu} \cdot {\bf d}^{*}_{\kappa})^{2} 
  \big] \nonumber
\end{eqnarray}
up to fourth order,
by analogy to He-3 \cite{Vollhardt90Taylor}.
Here $\nu, \kappa$ stand for orbital pairs $a/b$,
while the (real) quartic mixing parameters obey $C^{(i)}_{\nu
  \kappa}=C^{(i)}_{\kappa \nu}$
and the asymmetry between in-plane and out-of-plane orbitals
due to \emph{e.g.} inter-orbital hopping is reflected in 
distinct coefficients ($A_{yz/xz} \neq A_{yz/xy} = A_{xz/xy}$, etc.).
This form is dictated by gauge symmetry,
SU(2) spin rotational symmetry, time reversal symmetry 
and the underlying lattice symmetries,
and shows that the $C^{(3)}_{\nu \kappa}$ and $C^{(4)}_{\nu \kappa}$ terms
are sensitive to the relative orientation of the $d$-vectors, whereas
the $C^{(1)}_{\nu \kappa}$ and $C^{(5)}_{\nu \kappa}$ contributions
additionally depend on their relative complex phases.

However, once SO coupling is included,
 ${\bf d}_{xz/xy}$, ${\bf d}_{yz/xy}$, and ${\bf d}_{yz/xz}$ are pinned
along $x$, $y$, and $z$ directions, respectively, 
as shown in fig. 1.
Inversion/time reversal symmetry on the other hand is still preserved 
and reflected in the degeneracy of the orientations/chiralities 
$\{{\bf d}_{xz/xy}, {\bf d}_{yz/xy}, {\bf d}_{yz/xz}\}$
and $\{-{\bf d}_{xz/xy}, -{\bf d}_{yz/xy}, -{\bf d}_{yz/xz}\}$.
The pinning of the $d$-vectors occurs due to additional terms in the 
free energy such as 
$\sim a^{(1)} |d^{z}_{yz/xz}|^{2} 
+ a^{(2)} \big[|d^{z}_{yz/xy}|^2 + |d^{z}_{xz/xy}|^2\big]
+ b^{(1)} \big[|d^{x}_{yz/xz}|^{2} + |d^{y}_{yz/xz}|^{2}\big]
+ b^{(2)} \big[|d^{x}_{xz/xy}|^{2} + |d^{y}_{yz/xy}|^{2}\big]
+ c^{(1)} \big[d^{x}_{yz/xy} (d^{y}_{xz/xy})^{*} +  d^{y}_{yz/xy}
(d^{x}_{xz/xy})^{*} 
+ \text{c.c.}\big] + \cdots$,
where the expansion parameters depend on the SO coupling strength,
naively suggesting that $a^{(1)}, a^{(2)} < b^{(1)}, b^{(2)},
c^{(1)}$, etc. 
\footnote{Analyzing the energetics of a corresponding two orbital model 
one can indeed show that SO interaction tends to stabilize
\emph{e.g.} the $d^{z}_{yz/xz}$-component over $d^{x}_{yz/xz}$ or $d^{y}_{yz/xz}$.}
SO interaction furthermore leads to a linear coupling between a particular 
component of (inter-orbital) spin-triplet pairing and
(intra-orbital) spin-singlet pairing.  
For example, writing SO coupling between $yz$ and $xz$ orbitals in the form of 
$-i \lambda [\hat{\sigma^{z}}]_{\sigma \sigma'} 
(c^{yz \dagger}_{{\bf k} \sigma} c^{xz}_{{\bf k} \sigma'}
- c^{xz \dagger}_{{\bf k} \sigma} c^{yz}_{{\bf k} \sigma'})$
the following linear coupling is allowed in the 
GL free energy: 
\begin{eqnarray}
  \label{eq:GLEnergy}
  &-i& \lambda [\hat{\sigma}^{z}]_{\sigma \sigma'} 
  \langle c^{yz \dagger}_{{\bf k} \sigma} c^{xz}_{{\bf k} \sigma'}
  - c^{xz \dagger}_{{\bf k} \sigma} c^{yz}_{{\bf k} \sigma'} \rangle \\
  &\times& [i \hat{\sigma}^{y} \hat{\sigma}^{z}]_{\sigma \sigma'}  
  \langle 
  c^{yz}_{{\bf k} \sigma} c^{xz}_{-{\bf k} \sigma'} 
  - c^{xz}_{{\bf k} \sigma} c^{yz}_{-{\bf k} \sigma'}
  \rangle \nonumber \\
  &\times&  
  \left( [i \hat{\sigma}^y]_{\sigma \sigma'} 
  \langle c^{yz \dagger}_{{\bf k} \sigma} c^{yz \dagger}_{-{\bf k} \sigma'} \rangle
  + [i \hat{\sigma}^y]_{\sigma \sigma'} 
  \langle c^{xz \dagger}_{{\bf k} \sigma} c^{xz \dagger}_{-{\bf k} \sigma'} \rangle
  \right)
  \nonumber \\
  &\rightarrow&  i \lambda d^{z}_{yz/xz} \left(\Delta^{s}_{yz/yz} + \Delta^{s}_{xz/xz} \right)^{*} + \text{c.c.}
\end{eqnarray} 
Note that ${\bf d}_{yz/xz}$ prefers the $z$-direction 
by coupling to spin-singlet pairing with a relative
phase difference of $\pm \pi/2$ depending on the sign of $\lambda$.
This is consistent with our findings below that 
the spin-triplet order parameters are purely real 
while the spin-singlet amplitudes are purely imaginary. 
A similar analysis can be carried out for $d^{x}_{xz/xy}$ and $d^{y}_{yz/xy}$.
The overall order parameter for $yz$ and $xz$ orbitals then is
$d^z_{xz/yz} + i (\Delta^s_{xz/xz} + \Delta^s_{yz/yz})$.
Since the relative phase between the orbital-triplet 
spin-singlet and the orbital-singlet spin-triplet
order parameters is fixed, there should be a collective mode
representing a resonance of supercurrent flow between the 
coupled order parameters with an energy scale
of order 
$\sim \sqrt{|d^{z}_{a/b}|^2+|\Delta^{s}_{a/a}|^2 + |\Delta^{s}_{b/b}|^{2}}$.

Note that the above result is fundamentally different 
from similar two orbital models,
which lead to a single orbital-singlet spin-triplet $d$-vector
\cite{Dai08PRL,Spalek01PRB,Werner03PRB}.
The present model is also distinguished from other models where
the momentum dependence in the band pairing usually 
originates from  nonlocal momentum dependent interactions \cite{Ng00EPL},
whereas here it arises from  spin and orbital mixing in 
the Bloch bands as described next.

\section{Momentum-dependent pairing in the Bloch bands}
\begin{figure}[t!]
  \centering
  \includegraphics*[width=0.8\linewidth, clip]{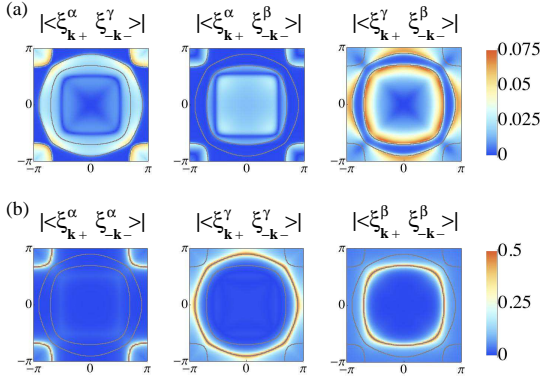}
  \caption{(Color online) 
    Momentum-resolved pairing amplitudes in the Bloch band basis 
    for $3J-U =0.9 $ and $\lambda = 0.15$.
    Panel (a) and (b) represent inter- and intra-band pairing, 
    respectively. The grey lines indicate the $\beta$, $\gamma$, and $\alpha$
    FS sheets (from inside to outside).
    Note that pairing from Hund's coupling preferentially involves
    electronic states near the FS sheets and that
    the intra-band pairing amplitudes are about one order of 
    magnitude larger than inter-band pairing amplitudes.
    \label{fig:BandPairing_l0.3_J0.9}}
\end{figure}

Despite having uniform pairing amplitudes 
${\bf d}_{yz/xz}, {\bf d}_{yz/xy}, {\bf d}_{xz/xy}, \Delta^{s}_{yz/yz}, \dots$
the corresponding inter- and intra-band pairings 
in the Bloch band basis (now carrying band and pseudospin
quantum numbers -- $\eta, \rho = \alpha, \beta, \gamma$
and $s=\pm$) acquire a strong momentum dependence
due to the mixing of orbitals through hopping and SO coupling.
To understand how the above local pairing in the orbital and spin basis
corresponds to pairing in the Bloch band basis,
let us introduce the kinetic Hamiltonian. 
The most generic kinetic Hamiltonian for t$_{\text{2g}}$ orbitals
in a single layer perovskite structure has the form
\begin{eqnarray}
  \label{eq:HKin}
  \! H_{\text{kin}} \! + H_{\text{SO}} \! \! &=& \!\!\! \sum_{{\bf k}, \sigma}
  \!
  C^{\dagger}_{{\bf k} \sigma} \!\!
  \begin{pmatrix}
    \varepsilon^{yz}_{\bf k}  & \varepsilon^{1d}_{\bf k} + i \lambda 
    & - \lambda  \\
    \varepsilon^{1d}_{\bf k} -i \lambda  & \varepsilon^{xz}_{\bf k} & i\lambda  \\
    -\lambda  & -i \lambda  & \varepsilon^{xy}_{\bf k} \\
  \end{pmatrix}
  \!\!
  C_{{\bf k} \sigma},
\end{eqnarray}
where $C^{\dagger}_{{\bf k} \sigma} 
= (c^{yz \dagger}_{{\bf k} \sigma}, c^{xz \dagger}_{{\bf k} \sigma}, 
c^{xy \dagger}_{{\bf k} -\sigma})$
and the dispersions are 
$\varepsilon^{yz/xz}_{\bf k} = - 2 t_{1} \text{cos}k_{y/x} - 2 t_{2} \text{cos}k_{x/y}
- \mu_{1}$,
$\varepsilon^{xy}_{\bf k} = - 2 t_{3} 
\big(\text{cos}k_{x} + \text{cos}k_{y} \big) 
- 4 t_{4} \text{cos}k_{x} \text{cos}k_{y} 
  %&& - 2 t_{5} \big( \text{cos}2k_{x} + \text{cos}2k_{y} \big) 
- \mu_{2}$, 
and
$\varepsilon^{1d}_{\bf k} = -4 t_{5} \text{sin}k_{x} \text{sin}k_{y}$.
For the MF calculation below we have chosen the parameters
$t_{1} = 0.5$, $t_{2} = 0.05$, $t_{3} = 0.5$, $t_{4} = 0.2$, $t_{5}=0.05$,
$\mu_{1}=0.55$, and $\mu_{2}=0.65$
(all energies here and in the following are expressed in units of $2t_{1}=1.0$).
The underlying FS obtained from diagonalizing 
$H_{\text{kin}}$ with SO coupling strength $\lambda = 0.15$ is shown in 
fig. \ref{fig:BandPairing_l0.3_J0.9} along with momentum-dependent
band pairing amplitudes.
The FS agrees well with first principles calculations \cite{Pavarini06PRB} 
and the experimentally measured FS of Sr$_{2}$RuO$_{4}$
\cite{Bergemann00PRL,Damascelli00PRL,Haverkort08PRL},
consisting of three bands labelled $\alpha$, $\beta$, and $\gamma$.

In the presence of SO coupling 
the bands are mixtures of all three orbitals and different spins, \emph{e.g.} 
$\xi^{\eta}_{{\bf k} +} = \tilde{f}^{\eta}_{\bf k} c^{xz}_{{\bf k} \uparrow}
+ \tilde{g}^{\eta}_{\bf k} c^{yz}_{{\bf k} \uparrow} 
+ \tilde{h}^{\eta}_{\bf k} c^{xy}_{{\bf k} \downarrow}$
($\eta=\alpha, \beta, \gamma$).
Hence considering inter- and intra-band pairing amplitudes
in the band basis, it is clear
that the $x$- and $y$-components of the inter-band pseudospin-triplets 
such as  
$\langle \xi^{\eta}_{{\bf k} \pm}  \xi^{\rho}_{-{\bf k} \pm} \rangle$
vanish, since  
$\langle d^{xz}_{{\bf k} \uparrow} d^{yz}_{-{\bf k} \uparrow} \rangle $,
$\langle d^{xz}_{{\bf k} \uparrow} d^{xy}_{-{\bf k} \downarrow} \rangle$, 
and $\langle d^{yz}_{{\bf k} \uparrow} d^{xy}_{-{\bf k} \downarrow} \rangle$
amplitudes are zero (similarly for $\uparrow \leftrightarrow \downarrow$).
Thus only finite  
$z$-components of the three inter-band pseudospin-triplet $d$-vectors
and inter-band pseudospin-singlet order parameters 
(such as 
$\langle \xi^{\eta}_{{\bf k} +} \xi^{\rho}_{-{\bf k} -} 
\pm \xi^{\rho}_{{\bf k} +} \xi^{\eta}_{-{\bf k} -} \rangle$) 
can appear. 
Figure \ref{fig:BandPairing_l0.3_J0.9} reveals that 
intra-band pairing is strongest and sharply peaked around the FS
due to the mixing of all orbitals via SO interaction and 
inter-orbital hopping, and the ideal conditions for 
zero-momentum pairing. 
Inter-band pairing in contrast is about an order of magnitude weaker
and, in particular for $\langle \xi^{\gamma}_{{\bf k} +}
\xi^{\beta}_{-{\bf k} -} \rangle$, more spread out in momentum
space, marking Bloch band states that are energetically still close enough to the
FS to participate significantly in pairing. 
 
This analysis demonstrates that inter-orbital pairing arising from 
Hund's interaction leads to ${\bf k}$-dependent 
inter- and intra-band pairing in pseudospin-singlet and
and pseudospin-triplet (z component only) channels.
Furthermore, the pairing instability occurs simultaneously within and between 
all bands rather than in a single active band with 
superconductivity leaking into passive bands through, e.g., pair
hopping.
The role of intra-band spin-triplet pairing between $\alpha$
and $\beta$ bands in multi-orbital superconductors like
Sr$_{2}$RuO$_{4}$ has also been the focus of recent studies, 
where the inter-band order parameter,
however, breaks TRS \cite{Raghu10PRL} and an intrinsic anomalous Hall effect can
contribute significantly to a large TRS breaking signal in Kerr rotation
experiments \cite{Taylor11arXiv,Wysokinski11arXiv}.

\begin{figure}[t!]
  \centering
  \includegraphics*[width=1.0\linewidth, clip]{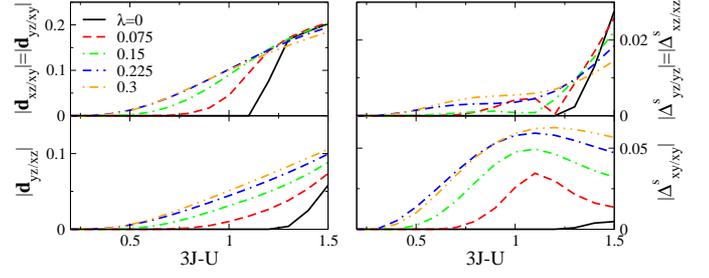}
  \caption{(Color online) MF solutions for different SO coupling strengths
    for the Sr$_{2}$RuO$_{4}$ based band structure.
    Orbital-singlet spin-triplet pairing ${\bf d}_{xz/xy}$, ${\bf d}_{yz/xy}$, 
    and ${\bf d}_{yz/xz}$ (purely real) induces 
    finite intra-orbital spin-singlet pairing
    $\Delta^{s}_{yz/yz}$, $\Delta^{s}_{xz/xz}$, and $\Delta^{s}_{xy/xy}$
    (purely imaginary).
    We also checked for induced inter-orbital 
    spin-singlet pairing amplitudes, which, however, vanish.
    \label{fig:RuthenateMFSol}}
\end{figure}

\section{Pairing transition, QP dispersion, and magnetic response}
For concreteness we study the effect of SO coupling on spin-triplet 
pairing originating 
from Hund's interaction, including the QP dispersion and the magnetic response.
As discussed in the previous sections the qualitative results are generic for 
SO coupled $t_{\text{2g}}$-bands (or p-orbital systems) and
can be applied to specific materials such as  
the single layer ruthenate \cite{Mackenzie03RMP,Kallin09JPhysCondensMatt}
and the Fe-pnictides \cite{Lee08PRB,Daghofer10PRB} using the 
appropriate band structure.

\begin{figure}[ttb]
  \centering
  \includegraphics*[width=0.8\linewidth, clip]{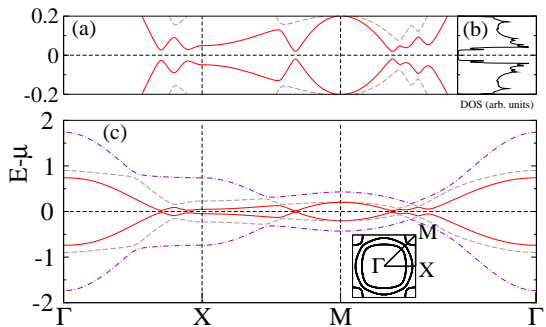}
  \caption{(Color online) QP bands for $3J-U=0.9$ and $\lambda=0.15$.
    Panel (a) is a magnification of panel (c) about the Fermi level,
    revealing the gaps opening up on the FS sheets.
    Panel (b) shows the DOS and the QP gap near the Fermi level.
    \label{fig:QPBands}}
\end{figure}

\begin{figure}[t!]
  \centering
  \includegraphics*[width=0.8\linewidth, clip]{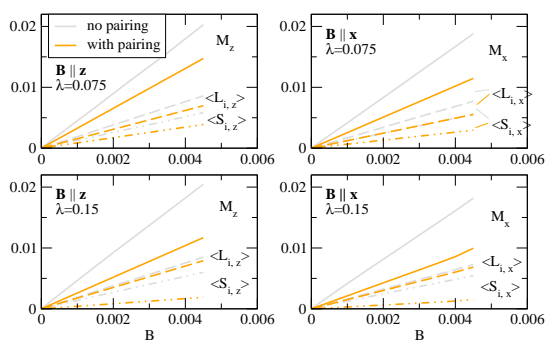}
  \caption{(Color online) Magnetization parallel to the applied magnetic 
    field ${\bf B}$ for $\lambda=0.075$ (top) and $0.15$ (bottom) and two field 
    orientations at $3J-U=0.9$.
    The solid lines represent total magnetization, dashed lines 
    stand for orbital contribution, and dash-dotted lines for
    spin magnetization. 
    For sake of comparison the magnetic response both in the presence 
    (orange) and in the absence (grey) of superconductivity is displayed.
    ($B$ is expressed in units of $2t_{1} = 1$.)
    \label{fig:Magnetization}} 
\end{figure}

Using the kinetic Hamiltonian of eq. (\ref{eq:HKin})
with a parameter choice mimicking the single layer ruthenate band structure,
the MF solutions for various $\lambda$
are displayed in fig. \ref{fig:RuthenateMFSol}.
As one can see, in the absence of SO interaction an orbital-singlet 
spin-triplet pairing instability develops at a
large coupling strength $3J-U \gtrsim 1.0$
for ${\bf d}_{xz/xy}$ and ${\bf d}_{yz/xy}$.
Although numerically difficult to resolve, we expect that
${\bf d}_{yz/xz}$ and the intra-orbital spin-singlet order parameters 
simultaneously become finite 
through quartic or  higher order
couplings in the Landau free energy expansion. 
While the magnitudes of the order parameters 
depend on the details of the band structure, a robust
feature is that finite SO coupling drastically reduces the critical 
pairing strength.
This reduction is mostly facilitated by the additional hybridization
provided by $H_{\text{SO}}$, which helps to overcome the momentum 
mismatch between orbitals/bands near the Fermi level.
On the other hand the same mechanism can have a slightly detrimental 
effect at larger $3J-U$, 
where the ideal inter-orbital pairing conditions along the diagonals 
are weakened by the additional hybridization.
One may also wonder if the Bogoliubov QP dispersions have anisotropic gaps.
The resulting QP bands are shown in fig. \ref{fig:QPBands} 
and are fully gapped with a fourfold symmetric gap modulation
in ${\bf k}$ space, even though the gap minima are tiny. 

Note that the present superconducting state does not 
break TRS.  
The magnetic response is a combination 
of paramagnetic (spin-triplet) and spin-singlet behaviours, 
with a slightly larger 
out-of-plane than in-plane total magnetic susceptibility as shown in 
fig. \ref{fig:Magnetization}, where ${\bf M}=\langle {\bf L}_{i} \rangle 
+ 2 \langle {\bf S}_{i}\rangle$
is the total magnetization including orbital and spin contributions
and $H_{\text{B}} = {\bf B} \cdot \sum_{i} ({\bf L}_{i} + 2 {\bf S}_{i})$
couples the orbital and spin degrees of freedom to the external field
${\bf B}$. 
Both orbital and spin expectation values are finite
with roughly equal contribution to the total magnetization. 
For comparison, the normal state magnetizations are also shown in fig. 
\ref{fig:Magnetization} and are larger than in the superconducting state,
as expected for a combination of spin-singlet and -triplet pairing 
in the presence of SO interaction. 
In particular, note that the spin magnetization changes drastically 
in the superconducting state with increasing $\lambda$. 
In general, the magnitude of the $d$-vectors, and thus the magnetic response,
can be modified by changing the size of the FS sheets.
For instance a larger overlap between yz and xy dominated portions of 
the FS would enhance ${\bf d}_{yz/xy}$ compared to ${\bf d}_{yz/xz}$ 
and ${\bf d}_{xz/xy}$.
The spin susceptibility then would be mostly dominated by 
${\bf d}_{yz/xy}$, a situation which may be facilitated 
by applying uniaxial pressure.

\section{Discussion and summary}
Given that we based our MF study on the Sr$_{2}$RuO$_{4}$
compound to illustrate the effect of SO interaction on pairing,
let us comment on the compatibility and the limitations
 of our results with what is known
about the superconducting state in Sr$_{2}$RuO$_{4}$
\cite{Mackenzie03RMP,Kallin09JPhysCondensMatt}.
Based on the QP gap variation along the FS sheets, one expects
that this modulation may also be reflected in orientation sensitive
specific heat measurements. 
Such magnetic field dependent specific heat measurements on Sr$_{2}$RuO$_{4}$
have indeed been carried out \cite{Deguchi04PRL,Deguchi04JPSJ},
but the interpretation of the experimental results is controversial, making a
link to our QP dispersion difficult. 
However, due to the nature of inter-band pairing, the superconducting 
state presented here is sensitive to any kind of impurities associated 
with inter-band scattering, which is consistent with the phenomena 
observed in Sr$_2$RuO$_4$.

Our result on the magnetization indicates that the 
spin-susceptibility is finite and different for in-plane and out-of-plane magnetic
field orientations in both the normal and the superconducting state,
as reported on Sr$_2$RuO$_4$.
Yet below $T_{\text{c}}$ the in-plane and out-of plane susceptibilities
decrease, which is in contrast to NMR Knight shift
measurements \cite{Ishida98Nature,Murakawa04PRL}, which revealed that 
a change in the spin-response across $T_{\text{c}}$ is absent for any
field orientation. 
This behaviour differs also from the response expected of a 
chiral $p+ip$ superconductor, where
the spin-susceptibility decreases for field directions
perpendicular to the $a$-$b$ plane but remains constant for parallel orientations. 
While the amount of change in the present model depends sensitively on 
the SO interaction strength, as shown in fig. \ref{fig:Magnetization}, 
the question also arises as to how orbital and spin  contributions were separated to obtain the Knight shift data
when SO interaction is significant.
Besides this, we note that the magnetic field effect on vortices will
be highly non-trivial as well, 
as it involves competition between various types of vortices 
including half-quantum vortices \cite{Salomaa85PRL,HYK00PRB}
in the presence of moderate SO coupling.

Finally, the lack of TRS breaking is compatible with the absence of chiral supercurrents
as observed in scanning Hall probe and scanning SQUID measurements
\cite{Bjoernsson05PRB,Kirtley07PRB}.
However, this contrasts with another proposal that the chiral states due to 
$p+ip$ pairing on $\alpha$ and $\beta$ bands
cancel each other leading to a topologically trivial superconductor
\cite{Raghu10PRL}.
It also contradicts Kerr rotation and $\mu$SR measurements which have been 
interpreted in favour of TRS breaking \cite{Luke98Nature,Xia06PRL}.
The issue as to whether TRS is broken or not is not 
 yet resolved in the experimental community.
While the current study supports a non-TRS breaking state, 
it can be modified by going beyond local interactions.  
A natural extension would be to include the effect 
of further neighbour ferromagnetic interactions such as those
discussed by Ng and Sigrist \cite{Ng00EPL}, which
could lead to a small admixture of odd parity pairing with broken 
TRS in addition to the pairing found here and 
which may be responsible for the broken TRS signatures
found in $\mu$SR and Kerr experiments \cite{Luke98Nature,Xia06PRL}.
Another possibility is a finite-momentum pairing state such as a
FFLO (Fulde-Ferrell-Larkin-Ovchinnikov) state \cite{Fulde64PR,Larkin65JETP}.
It is plausible that a FFLO state between different bands can be stabilized
over the inter-band pseudospin-triplet pairing. 
These studies, and more definite predictions for Sr$_2$RuO$_4$ or other
specific materials, however, go beyond the scope of the current 9 
complex order parameter minimization and require more detailed work.

In summary, we studied the combined effect of Hund's 
and SO coupling on  t$_{\text{2g}}$ orbital systems.  
Three orbital-singlet spin-triplet pairings 
were found to form an orthogonal $d$-vector triad.
A linear coupling between even-parity inter-orbital spin-triplet and 
even-parity intra-orbital spin-singlet pairings 
was allowed due to SO interaction, determining the orientation of the
three d-vectors and giving rise to a relative phase difference 
of $\pi/2$ between spin-singlet and spin-triplet order parameters.
We also showed that inter-orbital spin-triplet pairing in the orbital basis 
corresponds to ever-parity inter- and intra-band 
pairing in the Bloch band basis, 
and discussed how the pairing strength varies within the Bloch bands.
We further found that SO coupling assists Hund's coupling driven pairing,  
which generally leads to an anisotropic QP gap
and an orbital dependent magnetic response.

\vspace{-0.25cm}

\acknowledgments

\vspace{-0.25cm}

We thank S.~R. Julian, A. Paramekanti,
Y.-J. Kim, K.~S. Burch, and C. Kallin
for useful discussions.
HYK thanks the hospitality of the MPI-PKS, 
Dresden, Germany where a part of this work was carried out.
This work was supported by the NSERC of Canada and Canada Research Chair.


\vspace{-0.2cm}





\begin{thebibliography}{0}
\bibitem{Grewe91North} 
  \Name{Grewe N. \and Steglich F.}
  \Book{Handbook on the Physics and Chemistry of Rare Earths}
  \Vol{14}
  \Publ{North-Holland, Amsterdam}
  \Year{1991}
  Chapt. ``Heavy Fermions''.
\bibitem{Sigrist91RMP} 
  \Name{Sigrist M. \and Ueda K.}
  \Review{Rev. Mod. Phys.} \Vol{63} \Year{1991} \Page{239}.
\bibitem{Powell10arXiv} 
  \Name{Powell B.~J. \and McKenzie R.~H.}   
  \Review{Rep. Prog. Phys.} \Vol{74} \Year{2010} \Page{10301}.
\bibitem{Kamihara08JACS} 
  \Name{Kamihara Y.} 
  \Review{ J. Am. Chem. Soc.} \Vol{130} \Year{2008} \Page{3296}.
\bibitem{Mackenzie03RMP} 
  \Name{Mackenzie A.~P. \and Maeno Y.}
  \Review{Rev. Mod. Phys.} \Vol{75} \Year{2003} \Page{657} \& references therein. 
\bibitem{Kallin09JPhysCondensMatt} 
  \Name{Kallin C. \and Berlinsky J.}
  \Review{J. Phys.: Condens. Matter} \Vol{21} \Year{2009}
  \Page{164210} 
  \& references therein.
\bibitem{Lee08PRB} 
  \Name{Lee P.~A. \and Wen X.~G.}
  \Review{Phys. Rev. B} \Vol{78} \Year{2008} \Page{144517}.
\bibitem{Yang09PRB} 
  \Name{Yang W.~L. {\it et al.}} 
  \Review{Phys. Rev. B} \Vol{80} \Year{2009} \Page{014508}. 
\bibitem{Fazekas99World} 
  \Name{Fazekas P.}
  \Book{Lecture Notes on Electron Correlation and Magnetism}
  \Publ{World Scientific, Singapore} \Year{1999}.
\bibitem{Kuriyama10APL} 
  \Name{Kuriyama H. {\it et al.}}   
  \Review{Appl. Phys. Lett.} \Vol{96} \Year{2010} \Page{182103}.
\bibitem{Liebsch00PRL}  
  \Name{Liebsch A. \and Lichtenstein A.}
  \Review{Phys. Rev. Lett.} \Vol{84} \Year{2000} \Page{1591}.
\bibitem{Pchelkina07PRB} 
  \Name{Pchelkina Z.~V. {\it et al.}}
  \Review{Phys. Rev. B} \Vol{75} \Year{2007} \Page{035122}.
\bibitem{Mravlje11PRL} 
  \Name{Mravlje J. {\it et al.}}
  \Review{Phys. Rev. Lett.} \Vol{106} \Year{2011} \Page{096401}.
\bibitem{Pavarini06PRB} 
  \Name{Pavarini E. \and Mazin I. I.}
  \Review{Phys. Rev. B} \Vol{74} \Year{2006} \Page{035115}.
\bibitem{Malvestuto11PRB} 
   \Name{Malvestuto M. {\it et al.}}
   \Review{Phys. Rev. B} \Vol{83} \Year{2011} \Page{165121}.
\bibitem{Rozbicki11JPhysCondensMatter}  
   \Name{Rozbicki E.~J.,  Annett J.~F., Souquet J.~R. \and Mackenzie
     A.~P.}
   \Review{J. Phys.: Condens Matter} \Vol{23} \Year{2011}
   \Page{094201}.
\bibitem{Haverkort08PRL} 
  \Name{Haverkort M.~W. {\it et al.}}
  \Review{Phys. Rev. Lett.} \Vol{101} \Year{2008} \Page{026406}.
\bibitem{Ng00EPL} 
  \Name{Ng K.\,K. \and Sigrist M.}  
  \Review{Europhys. Lett.} \Vol{49} \Year{2000} \Page{473}.
\bibitem{Spalek01PRB} 
  \Name{Spalek J.}   
  \Review{Phys. Rev. B} \Vol{63} \Year{2001} \Page{104513}.
\bibitem{Han04PRB} 
  \Name{Han J.~E.}
  \Review{Phys. Rev. B} \Vol{70} \Year{2004} \Page{054513}.
\bibitem{Dai08PRL} 
  \Name{Dai X., Fang Z., Zhou Y. \and Zhang F.~C.}
  \Review{Phys. Rev. Lett.} \Vol{101} \Year{2008} \Page{057008}.
\bibitem{Leggett75RMP} 
  \Name{Leggett A.~J.}   
  \Review{Rev. Mod. Phys.} \Vol{47} \Year{1975} \Page{331}.
\bibitem{Vollhardt90Taylor} 
  \Name{Vollhardt D. \and W\"olfle P.}
  \Book{The Superfluid Phases of Helium 3}
  \Publ{Taylor \& Francis, London}
  \Year{1990},
  Chap. 5.
\bibitem{Werner03PRB} 
  \Name{Werner R.}   
  \Review{Phys. Rev. B} \Vol{67} \Year{2003} \Page{014505}.
\bibitem{Bergemann00PRL} 
  \Name{Bergemann C. {\it et al.}}
  \Review{Phys. Rev. Lett.} \Vol{84} \Year{2000} \Page{2662}.
\bibitem{Damascelli00PRL} 
  \Name{Damascelli A. {\it et al.}}
  \Review{Phys. Rev. Lett.} \Vol{85} \Year{2000} \Page{5194}.
\bibitem{Raghu10PRL} 
  \Name{Raghu S., Kapitulnik A. \and Kivelson S.~A.}
  \Review{Phys. Rev. Lett.} \Vol{105} \Year{2010} \Page{136401}.   
\bibitem{Taylor11arXiv} 
  \Name{Taylor E. \and Kallin C.}   
  \Review{Phys. Rev. Lett.} \Vol{108} \Year{2012} \Page{157001}.
\bibitem{Wysokinski11arXiv} 
  \Name{Wysokinski K.~I., Annett J.~F. \and Gy\"orffy B.~L.}   
  \Review{Phys. Rev. Lett.} \Vol{108} \Year{2012} \Page{077004}.
\bibitem{Daghofer10PRB} 
  \Name{Daghofer M., Nicholson A., Moreo A. \and Dagotto E.}
  \Review{Phys. Rev. B} \Vol{81} \Year{2010} \Page{014511}. 
\bibitem{Deguchi04JPSJ} 
  \Name{Deguchi K., Mao Z.~Q. \and Maeno Y.}   
  \Review{J. Phys. Soc. Jpn.} \Vol{73} \Year{2004} \Page{1313}.
\bibitem{Deguchi04PRL} 
  \Name{Deguchi K., Mao Z.~Q., Yaguchi H., \and Maeno Y.}
  \Review{Phys. Rev. Lett.} \Vol{92} \Year{2004} \Page{047002}.
\bibitem{Ishida98Nature} 
  \Name{Ishida K. {\it et al.}}
  \Review{Nature} \Vol{396} \Year{1998} \Page{658}.
\bibitem{Murakawa04PRL} 
  \Name{Murakawa H. {\it et al.}}
  \Review{Phys. Rev. Lett.} \Vol{93} \Year{2004} \Page{167004}.
\bibitem{Salomaa85PRL} 
  \Name{Salomaa M.~M. \and Volovik G.~E.}   
  \Review{Phys. Rev. Lett.} \Vol{55} \Year{1985} \Page{1184}.
\bibitem{HYK00PRB} 
  \Name{Kee H.-Y., Kim Y.~B. \and Maki K.}   
  \Review{Phys. Rev. B} \Vol{62} \Year{2000} \Page{R9275}.
\bibitem{Bjoernsson05PRB} 
  \Name{Bj\"ornsson P.~G., Maeno Y., Huber M.~E. \and 
    Moler K.~A.} 
  \Review{Phys. Rev. B} \Vol{72} \Year{2005} \Page{012504}.
\bibitem{Kirtley07PRB} 
  \Name{Kirtley J.~R. {\it et al.}} 
  \Review{Phys. Rev. B} \Vol{76} \Year{2007} \Page{014526}.
\bibitem{Luke98Nature} 
  \Name{Luke G.~M. {\it et al.}}
  \Review{Nature} \Vol{396} \Year{1996} \Page{658}.
\bibitem{Xia06PRL} 
  \Name{Xia J. {\it et al.}}
  \Review{Phys. Rev. Lett.} \Vol{97} \Year{2006} \Page{167002}.
\bibitem{Fulde64PR} 
  \Name{Fulde P. \and Ferrell R.~A.}   
  \Review{Phys. Rev.} \Vol{135} \Year{1964} \Page{A550}.
\bibitem{Larkin65JETP} 
  \Name{Larkin A.~I. \and Ovchinnikov Y.~N.}
  \Review{JETP} \Vol{20} \Year{1965} \Page{762}.
\end{thebibliography}
\end{document}